\documentclass{webofc}
\usepackage{hyperref}
\usepackage{mathtools,siunitx,booktabs,braket,slashed}
\usepackage{tikz}
\usepackage[export]{adjustbox} %includes package graphicx

\usepackage[percent]{overpic}

\makeatletter
\g@addto@macro\bfseries{\boldmath}
\g@addto@macro\normalfont{\unboldmath}
\makeatother

\DeclarePairedDelimiter\abs{\lvert}{\rvert}

\newcommand\amuhlbl{a_\mu^{\text{HLbL}}}
\newcommand\amulbl{a_\mu^{\text{LbL}}}
\newcommand\wtg[1]{\bar{\mathfrak{g}}^{(#1)}}
\newcommand\wtl[1]{\bar{\mathfrak{l}}^{(#1)}}

\newcommand\etal{\emph{et al.}}

\usepackage{graphicx}
\usepackage{epsfig}
\usepackage{rotating}
\usepackage{bm}
\newcommand{\be}{\begin{equation}}
\newcommand{\ee}{\end{equation}}
\newcommand{\ba}{\begin{eqnarray}}
\newcommand{\ea}{\end{eqnarray}}
\newcommand{\bi}{\begin{itemize}}
\newcommand{\ei}{\end{itemize}}

\newcommand{\<}{\langle} 
\renewcommand{\>}{\rangle}

\newcommand{\la}{\label}

\title{Exploratory studies for the position-space approach to hadronic 
light-by-light scattering in the muon $g-2$}
\author{
   \firstname{Nils} \lastname{Asmussen} \inst{1} 
   \and
   \firstname{Antoine} \lastname{G\'erardin} \inst{1}
   \and
   \firstname{Harvey B.} \lastname{Meyer} \inst{1,2}
   \and
   \firstname{Andreas} \lastname{Nyffeler} \inst{1}
   \\
   \mbox{E-mail: \{asmussen,gerardin,meyerh,nyffeler\}@uni-mainz.de}
}

\institute{
   PRISMA Cluster of Excellence \& Institut f\"ur Kernphysik, \\
   Johannes Gutenberg-Universit\"at Mainz, 55099 Mainz, Germany 
\and
  Helmholtz Institute Mainz, Johannes Gutenberg-Universit\"at Mainz, 55099 Mainz, Germany}

\abstract{
The well-known discrepancy in the muon $g-2$ between experiment and theory 
   demands further theory investigations in view of the upcoming new 
   experiments.  One of the leading uncertainties lies in the hadronic 
   light-by-light scattering contribution (HLbL), that we address with our 
   position-space approach. We focus on exploratory studies of the pion-pole 
   contribution in a simple model and the fermion loop without gluon exchanges 
   in the continuum and in infinite volume. These studies provide us with useful 
   information for our planned computation of HLbL in the muon $g-2$ using full 
   QCD.}

\begin{document}

\maketitle

\section{Introduction}
The anomalous magnetic moment of the muon \(a_\mu=\frac{g_\mu-2}{2}\) provides a 
high-precision test of the Standard Model. Current experiments and 
Standard-Model computations show a discrepancy of about three standard 
deviations; see Ref.~\cite{Jegerlehner:2017lbd} and references therein. This 
leads to the question whether this is a hint of new physics or just a 
statistical or systematic fluctuation from the exact value. To address this 
question, the uncertainty on this value has to be reduced. Experiments at 
Fermilab and at J-PARC plan to improve on the uncertainty by a factor of 
four\cite{hertzog:2015jru}.  The theoretical prediction ought to be improved in 
equal measure. The theoretical uncertainty is dominated by the hadronic vacuum 
polarization contribution (HVP) and the hadronic light-by-light contribution 
(HLbL).

Lattice QCD can provide a first-principle estimate of 
$\amuhlbl$~\cite{Blum:2014oka,Blum:2015gfa,Blum:2016lnc,Blum:2017cer,Asmussen:2015,Green:2015mva,Asmussen:2016lse}.
Other methods rely on models, because the HLbL is not fully related to any cross 
section, leading to large uncertainties. Dispersion relations allow one to use 
experimental data to reduce the uncertainties for the dominant contributions 
(\(\pi^0\,,\,\eta\,,\,\eta^\prime\,;\,\pi\pi\)), see Colangelo \etal{} 
\cite{Colangelo:2014dfa,Colangelo:2014pva,Colangelo:2015ama,Colangelo:2017qdm,Colangelo:2017fiz} 
and Pauk and Vanderhaeghen~\cite{Pauk:2014rfa}. Results from lattice QCD can be 
used as inputs to~\cite{Gerardin:2016cqj} or tests of~\cite{Green:2015sra} these 
dispersive approaches. More challenging is a full calculation of \(\amuhlbl\); 
in these proceedings, we report our progress towards such a calculation.

\section{Expression for $\amuhlbl$ in Euclidean position space}
\label{sec:method}
\begin{figure}
   \centerline{%
      \begin{overpic}[width=0.36\textwidth]{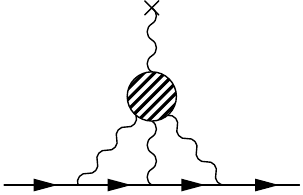}
         \put(39,24){\small $x$}
         \put(53,20){\small $y$}
         \put(59,26){\small $0$}
         \put(51,42){\small $z$}
      \end{overpic}
   }
   \caption{Hadronic light-by-light scattering in the $(g-2)_\mu$. The muon and the photon 
   propagators are contained in the QED kernel function~\(\bar{\mathcal 
   L}\). The blob denotes the QCD correlation function~\(i\hat\Pi\) to be 
   evaluated on the lattice.}
   \label{fig:muonhlbl}
\end{figure}
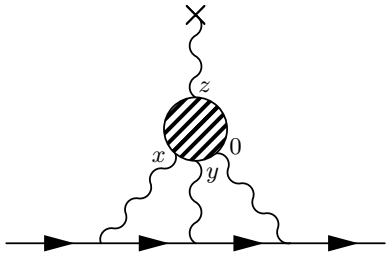

The HLbL can be split into a continuum, infinite volume QED kernel function 
\(\bar{\cal L}\) and a (Lattice) QCD four-point correlation function 
\(i\widehat\Pi\); see Fig.~\ref{fig:muonhlbl}.  The anomalous magnetic 
moment \(\amuhlbl\) can then be computed from our master formula:
\begin{align}
   \label{eq:master}
   \amuhlbl =& F_2(0) = \frac{m e^6}{3}  \underbrace{\int 
   d^4y}_{\hspace{-4.0em}=2\pi^2 \int_0^\infty d|y| |y|^3}
   \Big[
      \underbrace{\int 
      d^4x}_{\mathclap{\raisebox{-0.5em}{\scriptsize$\hspace{8em}=4\pi\int_0^\pi 
      d\beta\sin^2(\beta)\int_0^\infty d\abs{x}\abs{x}^3$}}}
      \underbrace{\bar{\cal L}_{[\rho,\sigma];\mu\nu\lambda}(x,y)}_{\rm QED}\;  
      \underbrace{i\widehat\Pi_{\rho;\mu\nu\lambda\sigma}(x,y)}_{\rm QCD}
   \Big],
   \\
   i\widehat \Pi_{\rho;\mu\nu\lambda\sigma}( x, y)  =& -\int d^4z\; z_\rho\, 
   \Big\<\,j_\mu(x)\,j_\nu(y)\,j_\sigma(z)\, j_\lambda(0)\Big\>.
\end{align}

For information on the derivation of the master formula, see our 
\emph{Lattice 2016} proceedings contribution~\cite{Asmussen:2016lse}. Note, that by treating 
the QED kernel function in infinite volume, we avoid introducing 
\(\frac{1}{L^2}\) finite-volume effects due to the photons. In the derivation we 
made the Lorentz covariance manifest, which allows us to reduce the 
eight-dimensional integral to an integral in three dimensions, as annotated in 
formula~\eqref{eq:master}. In a Lattice QCD computation of the fully connected 
diagrams, one would evaluate the four-dimensional \(x\) integral with the help 
of sequential propagators and only reduce the \(y\) integral to one dimension.  
The square brackets in formula~\eqref{eq:master} can be evaluated in one step 
for one value of $y$.  This makes it affordable to sample the integrand for the 
remaining one-dimensional integral
over \(y\).

The kernel function \(\bar{\mathcal L}\) is decomposed into tensors:
\begin{align}
\bar {\cal L}_{[\rho,\sigma];\mu\nu\lambda}(x,y) 
= \sum_{A={\rm I,II,III}} {\cal G}^A_{\delta[\rho\sigma]\mu\alpha\nu\beta\lambda} T^{(A)}_{\alpha\beta\delta}(x,y) ,
\end{align}
with e.\,g.\
\ba \la{eq:GI}
{\cal G}^{\rm I}_{\delta[\rho\sigma]\mu\alpha\nu\beta\lambda} &\equiv &
{\frac{1}{8}} {\rm Tr}\Big\{\Big( \gamma_\delta [\gamma_\rho,\gamma_\sigma] +
2 (\delta_{\delta\sigma}\gamma_\rho - \delta_{\delta\rho}\gamma_\sigma)\Big)
\gamma_\mu \gamma_\alpha \gamma_\nu \gamma_\beta \gamma_\lambda \Big\}.
\ea
The trace of the gamma matrices evaluates to sums of products of Kronecker 
deltas.

\begin{figure}
   \resizebox{\textwidth}{!}{
      \begin{tabular}{rrr}
         \includegraphics[valign=b]{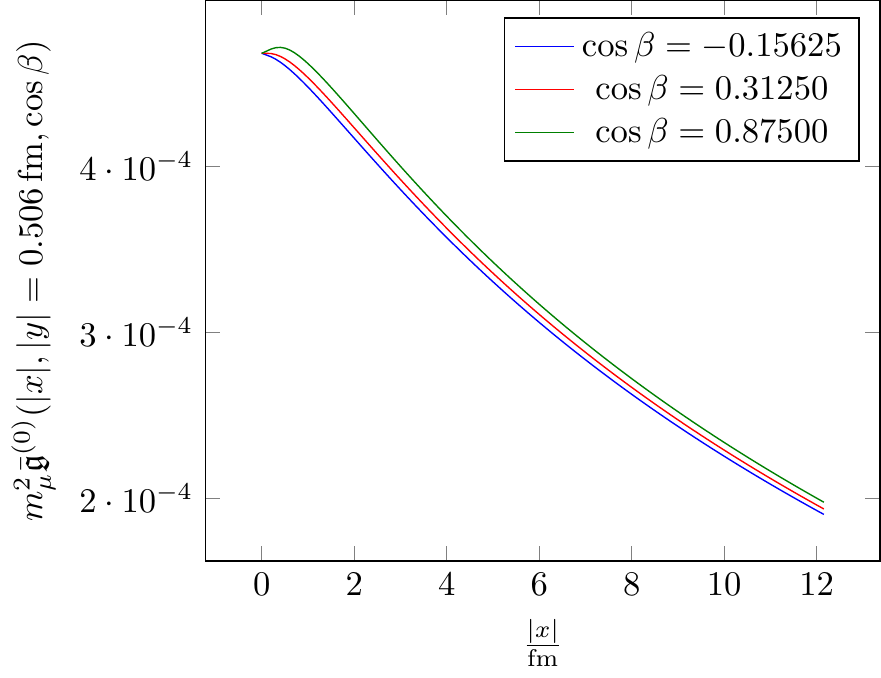} &
         \includegraphics[valign=b]{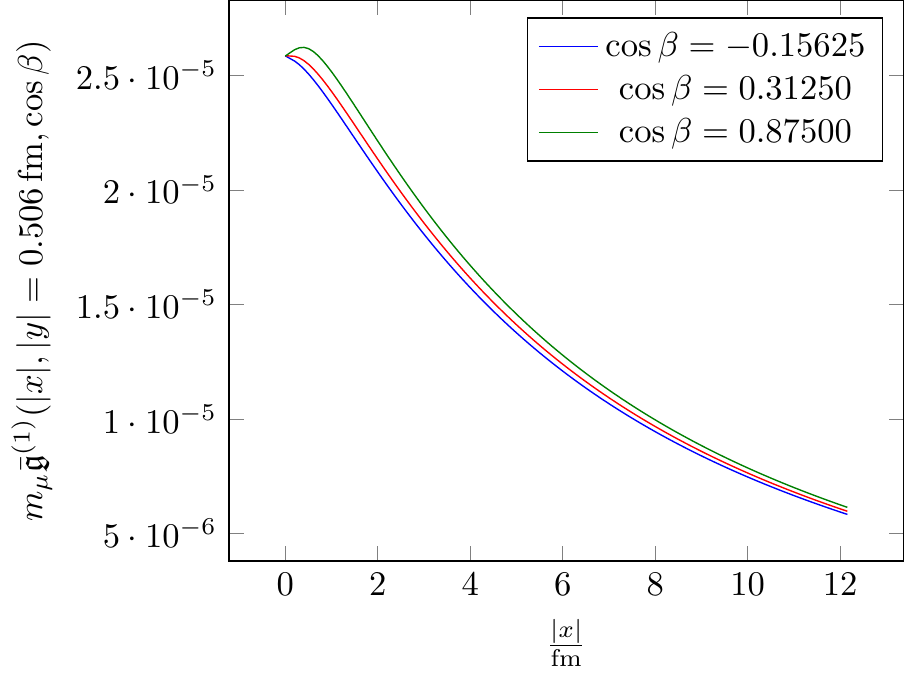} &
         \includegraphics[valign=b]{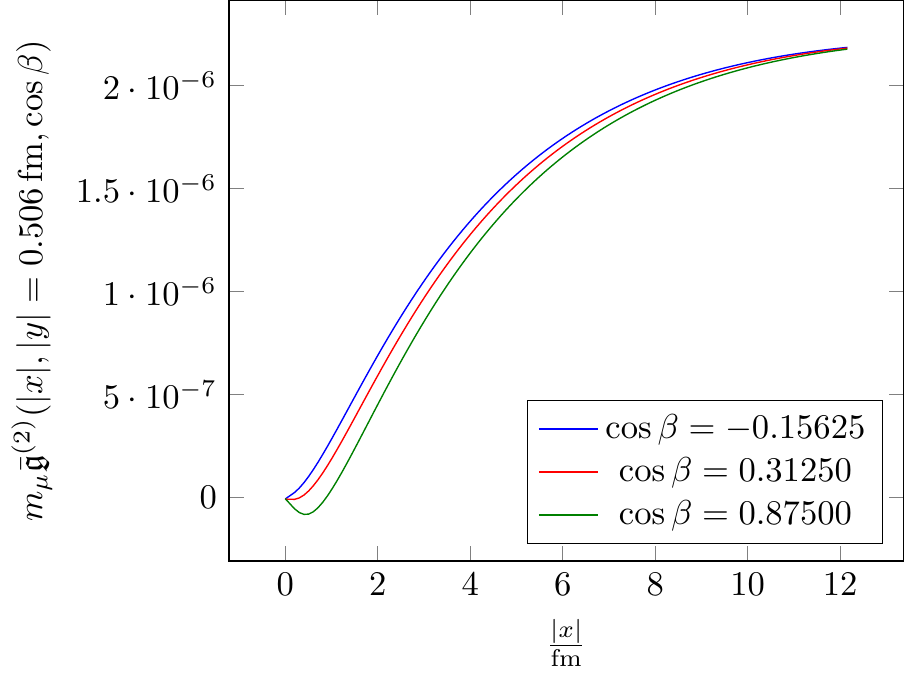}\\
         \includegraphics[valign=b]{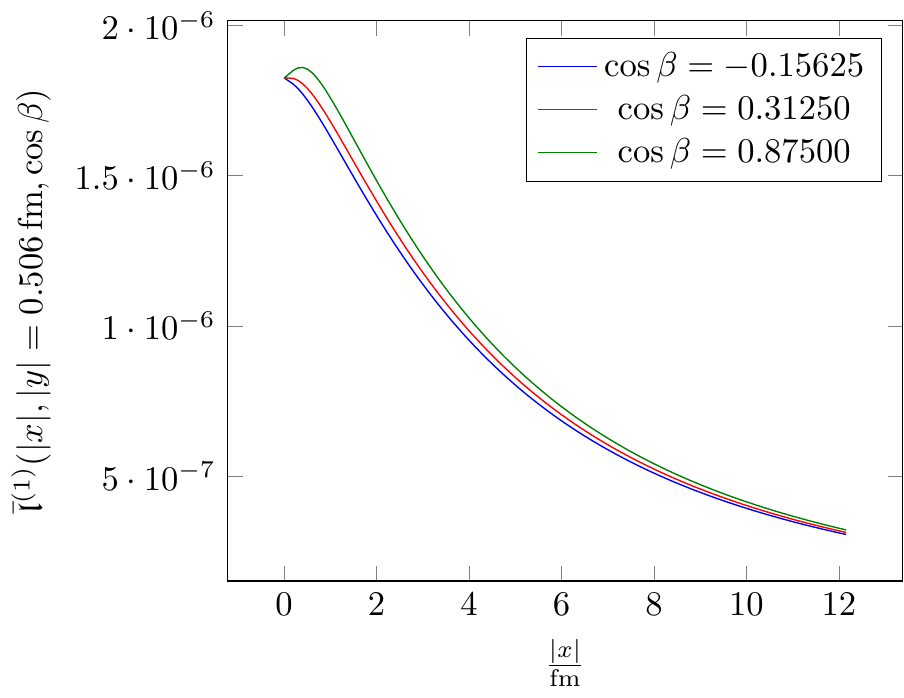} &
         \includegraphics[valign=b]{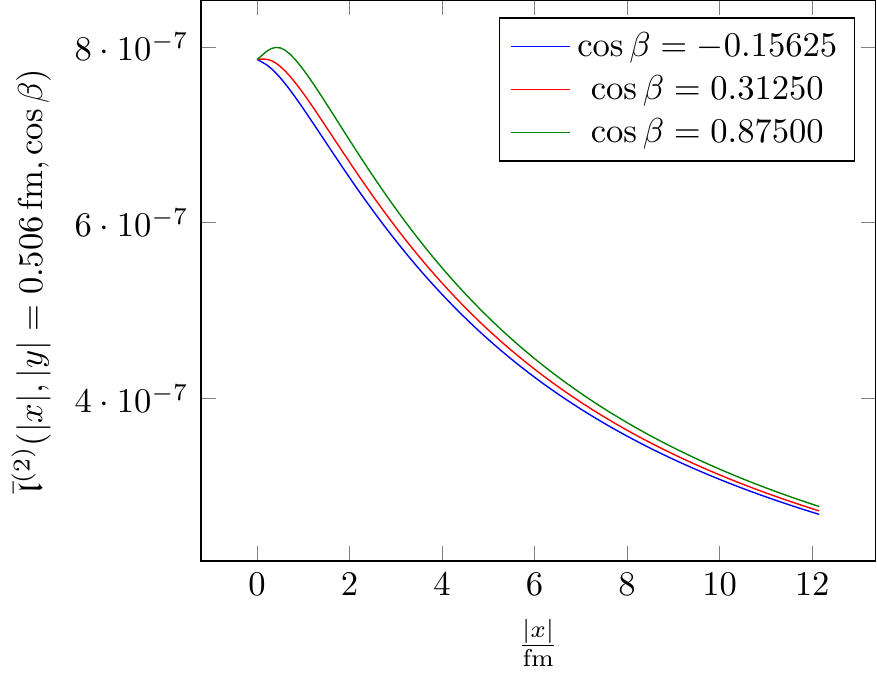} &
         \includegraphics[valign=b]{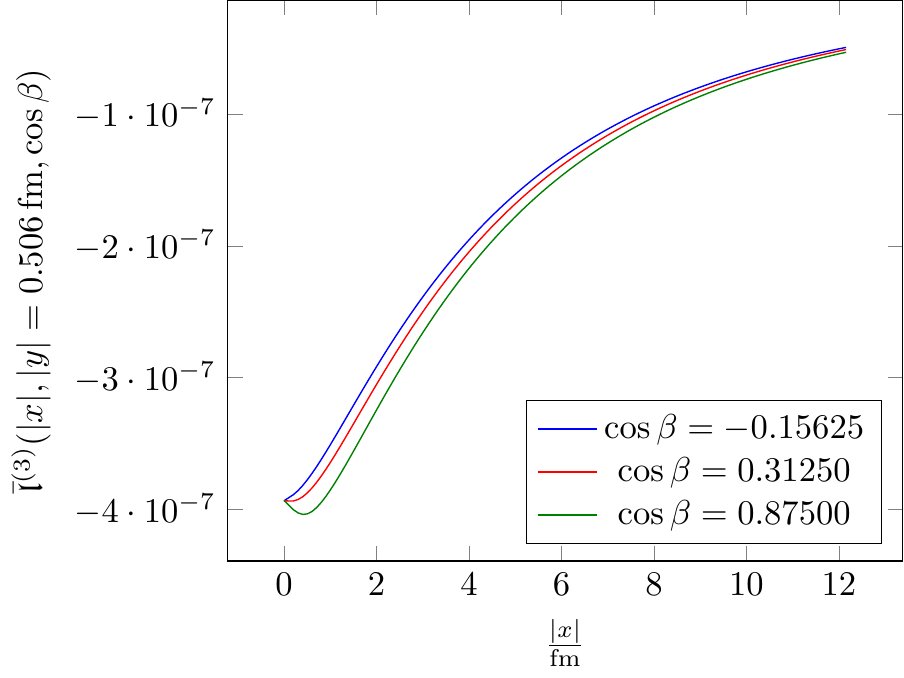}
      \end{tabular}
   }
   \caption{The $\abs x$ dependence of all six weight functions for $\abs 
   y=\SI{0.506}{fm}$ and three values of $\cos\beta$. Note that \(\wtg{0}\) 
   contains an arbitrary additive constant, because of a regulated 
   infrared divergence present before taking the derivatives in 
   Eqs.~(\ref{eq:TI}--\ref{eq:TIII}).}
   \label{fig:weights}
\end{figure}

The tensors \(T^{(A)}\) in turn are decomposed into a scalar~\(S\), 
vector~\(V\) and tensor~\(T\) contribution,
\ba
\la{eq:TI}
T^{({\rm I})}_{\alpha\beta\delta}(x,y) &=&   \partial^{(x)}_\alpha (\partial^{(x)}_\beta + \partial^{(y)}_\beta) 
V_\delta(x,y),
\\
\la{eq:TII}
T^{({\rm II})}_{\alpha\beta\delta}(x,y) &=& 
m \partial^{(x)}_\alpha 
\Big( T_{\beta\delta}(x,y) + \frac{1}{4}\delta_{\beta\delta} S(x,y)\Big),
\\
\la{eq:TIII}
T^{({\rm III})}_{\alpha\beta\delta}(x,y) &=&  m (\partial^{(x)}_\beta + \partial^{(y)}_\beta)
\Big( T_{\alpha\delta}(x,y) + \frac{1}{4}\delta_{\alpha\delta} S(x,y)\Big).
\ea
The latter are parametrized by the six weight functions 
\(\wtg{0,1,2}\), \(\wtl{1,2,3}\):
{\small%
   \begin{align}
   S(x,y) &= \wtg{0}(|x|,\hat x\cdot \hat y, |y|),  \phantom{\frac{1}{1}}
   \\
   V_\delta(x,y)
   &= x_\delta  \bar{\mathfrak{g}}^{(1)}(|x|,\hat x\cdot\hat y,|y|)
   + y_\delta  \bar{\mathfrak{g}}^{(2)}(|x|,\hat x\cdot\hat y,|y|),
   \\
    T_{\alpha\beta}(x,y) &= (x_\alpha x_\beta - 
      \frac{x^2}{4}\delta_{\alpha\beta})\; \bar{\mathfrak{l}}^{(1)}
   + (y_\alpha y_\beta - \frac{y^2}{4}\delta_{\alpha\beta})\; \bar{\mathfrak{l}}^{(2)}
   + (x_\alpha y_\beta + y_\alpha x_\beta - \frac{x\cdot y}{2}\delta_{\alpha\beta})\; \bar{\mathfrak{l}}^{(3)},
   \end{align}%
}%
where \(\hat x=\frac{x}{\abs x}\) and \(\hat y=\frac{y}{\abs y}\).
To evaluate the QED kernel $\bar{\cal L}_{[\rho,\sigma];\mu\nu\lambda}(x,y)$ we 
compute and store all six weight functions; the remaining 
operations to get the QED kernel are computationally inexpensive and 
it is convenient to perform them during the lattice computation.
Due to the Lorentz covariance, the six weight functions \(\wtg{0,1,2}\) and \(\wtl{1,2,3}\) 
are functions of the three parameters \(x^2\), \(y^2\) and \(x\cdot y\) only. Therefore, it is feasible to 
precompute and store them. Plots of all six weight functions are shown in Fig.~\ref{fig:weights}.

\section{Numerical tests}
To verify that the method and its implementation are correct, we computed the 
\(\pi^0\)-pole contribution in the vector-meson-dominance model as well as the 
lepton-loop contribution to \(\amulbl\) in QED. These results can be compared 
with the known values of these contributions.

\subsection{The $\pi^0$-pole contribution}
The first check is the $\pi^0$-pole contribution assuming a 
vector-meson-dominance transition form factor (parameters: $m_V$, $m_\pi$ and 
overall normalization),
\begin{align}\label{eq:Fpi0}
   {\cal F}(-q_1^2,-q_2^2) = \frac{c_\pi}{(q_1^2+m_V^2)(q_2^2+m_V^2)}, \qquad 
   c_\pi = -\frac{N_c m_V^4}{12\pi^2 F_\pi}.
\end{align}

We construct the correlation function for the \(\pi^0\)-pole contribution. The 
result reads
\begin{align}
   & i \widehat\Pi_{\rho;\mu\nu\lambda\sigma}(x,y)
   = \frac{c_\pi^2}{m_V^2(m_V^2-m_\pi^2)} \frac{\partial}{\partial 
   x_\alpha}\frac{\partial}{\partial y_\beta}
   \Big\{ 
    \epsilon_{\mu\nu\alpha\beta}\epsilon_{\sigma\lambda\rho\gamma}
   \Big(\frac{\partial}{\partial x_\gamma}+\frac{\partial}{\partial y_\gamma}\Big) 
   K_\pi(x,y)
   \nonumber\\ & + \epsilon_{\mu\lambda\alpha\beta} 
   \epsilon_{\nu\sigma\gamma\rho}
    \frac{\partial}{\partial y_\gamma} 
   K_\pi(y-x,y)
   + \epsilon_{\mu\sigma\alpha\rho}\epsilon_{\nu\lambda\beta\gamma} 
   \frac{\partial}{\partial x_\gamma}
   K_\pi(x,x-y)
    \Big\},
   \la{eq:PihatPi0Master}
\end{align}
where, with the massive propagator in position space \(G_m(x)\),
\begin{align}
   K_\pi(x,y) \equiv \int d^4u \Big(G_{m_\pi}(u) - G_{m_V}(u)\Big)
   G_{m_V}(x-u) G_{m_V}(y-u) = K_\pi(y,x).
   \la{eq:PiK}
\end{align}

\begin{figure}
   \includegraphics[width=0.49\textwidth]{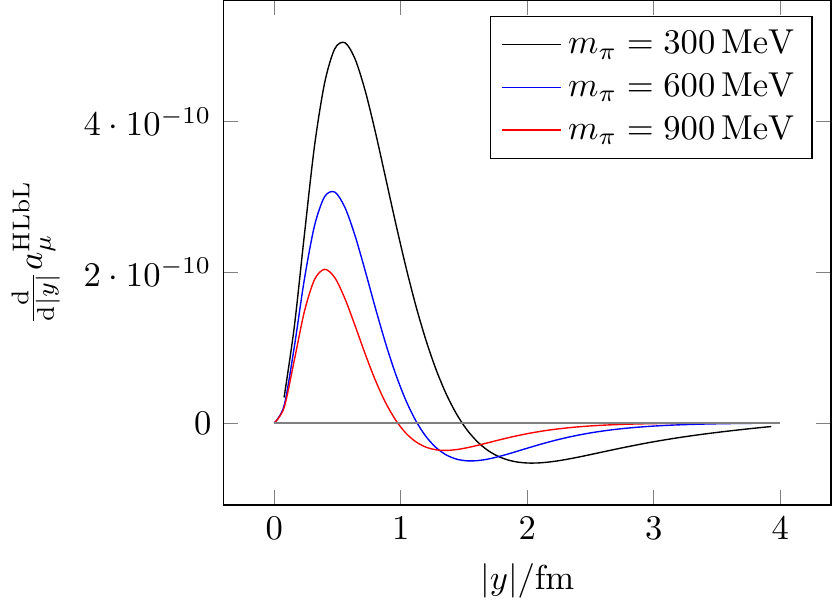}
   \includegraphics[width=0.49\textwidth]{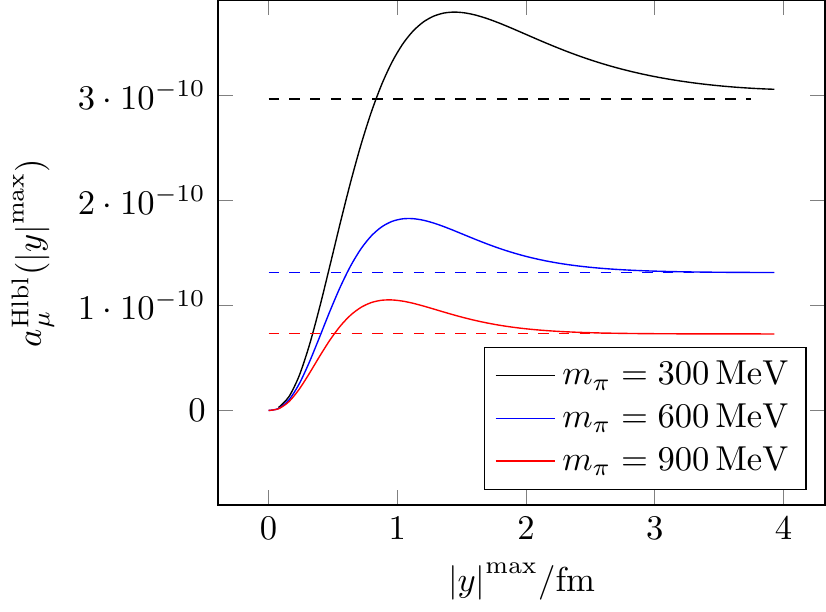}
   \caption{The \(\pi^0\)-pole contribution assuming a transition form factor 
   given by the vector-meson-dominance model.
   For a cutoff choosen to be \SI{4}{fm} in the \(\abs{x}\) direction, the left 
   plot shows the integrand of the \(\abs y\) integration of \(\amuhlbl\) and 
   the right plot shows the value of \(\amuhlbl\) for an upper
   integration limit \(\abs{y}^{\text{max}}\). The dashed line
   represents the result from the momentum-space integration.}
   \label{fig:pi0vmd}
\end{figure}

With the correlation function at hand, we can apply the techniques described in 
Sec.~\ref{sec:method}. The result is shown in Fig.~\ref{fig:pi0vmd}. In view of 
the exponential decay $\sim e^{- \tilde c m_\pi |y|}$ of the correlation 
function, the observed contribution to $\amuhlbl$ is remarkably long-range. This 
demands for large lattices at the order of $5-10\,$\si{fm} for the physical pion 
mass.

\subsection{The lepton loop contribution in QED}
The analytic result for the correlation function for a lepton loop with mass 
\(m_l\) is:
\ba\nonumber
i\widehat\Pi_{\rho;\mu\nu\lambda\sigma}(x,y) &=& \widehat\Pi^{(1)}_{\rho;\mu\nu\lambda\sigma}(x,y)
\\ && +\widehat\Pi^{(1)}_{\rho;\nu\lambda\mu\sigma}(y-x,-x) + x_\rho\, \Pi^{(r,1)}_{\nu\lambda\mu\sigma}(y-x,-x)
\nonumber  \\ && +  \widehat\Pi^{(1)}_{\rho;\lambda\nu\mu\sigma}(-x,y-x) + x_\rho\,\Pi^{(r,1)}_{\lambda\nu\mu\sigma}(-x,y-x).
\ea
It consists of the two functions \(\Pi^{(1)}\) and \(\Pi^{(r,1)}\). These 
functions are sums of products of Bessel functions and traces of gamma matrices.  
The gamma matrices evaluate to sums of products of Kronecker deltas. The 
integral in \(z\) has already been performed analytically and the evaluation 
boils down to computing the Bessel functions and evaluating the traces. The two 
functions read
\begin{align}
   \MoveEqLeft
   \Pi^{(r,1)}_{\mu\nu\lambda\sigma}(x,y)
   = 2\Big(\frac{m_l}{2\pi}\Big)^8 \Big[
   \nonumber\\&
   \frac{(-x_\alpha) (x-y)_\beta \; K_2(m_l|x|) K_2(m_l|x-y|)}{|x|^2 |x-y|^2 
   }\cdot 
   l_{\gamma\delta}(y) \cdot
   {\rm Tr}\{ \gamma_\alpha \gamma_\mu \gamma_\beta \gamma_\nu \gamma_\gamma 
   \gamma_\sigma \gamma_\delta \gamma_\lambda\}
   \nonumber\\&
   + \frac{ K_1(m_l|x|) K_1(m_l|x-y|)}{ |x| |x-y| } \cdot 
   p(|y|) \cdot  {\rm Tr}\{  \gamma_\mu  \gamma_\nu  \gamma_\sigma  \gamma_\lambda\}
   \nonumber\\&
   + \frac{(-x_\alpha) (x-y)_\beta \; K_2(m_l|x|) K_2(m_l|x-y|) }{ |x|^2 |x-y|^2 
   }\cdot
   p(|y|) \cdot {\rm Tr}\{ \gamma_\alpha \gamma_\mu \gamma_\beta \gamma_\nu  \gamma_\sigma \gamma_\lambda\}
   \nonumber\\&
   +  \frac{(-x_\alpha)  \; K_2(m_l|x|) K_1(m_l|x-y|)}{|x|^2 |x-y| }\cdot
   q_\gamma(y) \cdot   {\rm Tr}\{ \gamma_\alpha \gamma_\mu  \gamma_\nu \gamma_\gamma \gamma_\sigma \gamma_\lambda\}
   \nonumber\\&
    +  \frac{ (x-y)_\beta  \; K_1(m_l|x|) K_2(m_l|x-y|)}{|x| |x-y|^2 }\cdot
   q_\gamma(y) \cdot    {\rm Tr}\{  \gamma_\mu \gamma_\beta \gamma_\nu \gamma_\gamma \gamma_\sigma \gamma_\lambda\}
   \nonumber\\&
    +  \frac{(-x_\alpha) \; K_2(m_l|x|) K_1(m_l|x-y|)}{|x|^2 |x-y|}\cdot
   q_\delta(y) \cdot   {\rm Tr}\{ \gamma_\alpha \gamma_\mu  \gamma_\nu  \gamma_\sigma \gamma_\delta \gamma_\lambda\}
   \nonumber\\&
   +  \frac{(x-y)_\beta \; K_1(m_l|x|) K_2(m_l|x-y|)}{|x| |x-y|^2}\cdot 
   q_\delta(y) \cdot    {\rm Tr}\{  \gamma_\mu \gamma_\beta \gamma_\nu  \gamma_\sigma \gamma_\delta \gamma_\lambda\}
   \nonumber\\&
   + \frac{ K_1(m_l|x|) K_1(m_l|x-y|)}{|x| |x-y| }\cdot
   l_{\gamma\delta}(y) 
   \cdot   {\rm Tr}\{  \gamma_\mu  \gamma_\nu \gamma_\gamma \gamma_\sigma \gamma_\delta \gamma_\lambda\} \Big]\,,
\end{align}
and
\begin{align}\la{eq:d4zI1}
   \MoveEqLeft
   \widehat\Pi^{(1)}_{\rho;\mu\nu\lambda\sigma}(x,y)
   = 2\Big(\frac{m_l}{2\pi}\Big)^8 \Big[
   \nonumber\\&
    \frac{(-x_\alpha) (x-y)_\beta  K_2(m_l|x|) K_2(m_l|x-y|)}{|x|^2 
    |x-y|^2}\cdot   {f}_{\rho\delta\gamma}(y)
    \cdot {\rm Tr}\{ \gamma_\alpha \gamma_\mu \gamma_\beta \gamma_\nu 
    \gamma_\gamma \gamma_\sigma \gamma_\delta \gamma_\lambda\}
   \nonumber\\&
   + \frac{  K_1(m_l|x|) K_1(m_l|x-y|)}{|x| |x-y| }\cdot
   {f}_{\rho\delta\gamma}(y)\cdot
   {\rm Tr}\{  \gamma_\mu  \gamma_\nu \gamma_\gamma \gamma_\sigma \gamma_\delta 
   \gamma_\lambda\}
   \nonumber\\&
   + \frac{ K_1(m_l|x|) K_1(m_l|x-y|)}{ |x| |x-y| }\cdot\,g_\rho(y)\cdot
    {\rm Tr}\{  \gamma_\mu  \gamma_\nu  \gamma_\sigma  \gamma_\lambda\}
   \nonumber\\&
   +  \frac{(-x_\alpha) (x-y)_\beta \; K_2(m_l|x|) K_2(m_l|x-y|)}{ |x|^2 
   |x-y|^2}\cdot
   g_\rho(y)\cdot
   {\rm Tr}\{ \gamma_\alpha \gamma_\mu \gamma_\beta \gamma_\nu  \gamma_\sigma 
   \gamma_\lambda\}
   \nonumber\\&
   + \frac{(-x_\alpha) \; K_2(m_l|x|) K_1(m_l|x-y|)}{|x|^2 |x-y| }\cdot
   h_{\rho\gamma}(y)\cdot
   {\rm Tr}\{ \gamma_\alpha \gamma_\mu  \gamma_\nu \gamma_\gamma \gamma_\sigma 
   \gamma_\lambda\}
   \nonumber\\&
   + \frac{ (x-y)_\beta  \; K_1(m_l|x|) K_2(m_l|x-y|)}{|x| |x-y|^2 }\cdot
   h_{\rho\gamma}(y)\cdot
   {\rm Tr}\{  \gamma_\mu \gamma_\beta \gamma_\nu \gamma_\gamma \gamma_\sigma 
   \gamma_\lambda\}
   \nonumber\\&
   + \frac{(-x_\alpha) \; K_2(m_l|x|) K_1(m_l|x-y|)}{|x|^2 |x-y| }\cdot
   \hat f_{\rho\delta}(y)\cdot
   {\rm Tr}\{ \gamma_\alpha \gamma_\mu  \gamma_\nu  \gamma_\sigma \gamma_\delta 
   \gamma_\lambda\}
   \nonumber\\&
   +  \frac{(x-y)_\beta \; K_1(m_l|x|) K_2(m_l|x-y|)}{|x| |x-y|^2}\cdot
   \hat f_{\rho\delta}(y)\cdot
   {\rm Tr}\{  \gamma_\mu \gamma_\beta \gamma_\nu  \gamma_\sigma \gamma_\delta 
   \gamma_\lambda\}
   \Big]\,,
\end{align}
where
\begin{align}
   l_{\gamma\delta}(y) &= \frac{2\pi^2}{m_l^2}\Big(
      \hat y_\gamma \hat y_\delta\, K_2(m_l|y|)
      - \delta_{\gamma\delta}\, \frac{K_1(m_l|y|)}{m_l|y|}
   \Big),
   \\
   h_{\rho\gamma}(y) &= \frac{\pi^2}{m_l^3} \Big(
      \hat y_\gamma \hat y_\rho\, m_l|y|\,K_1(m_l|y|)
      - \delta_{\gamma\rho} K_0(m_l|y|)
   \Big),\\
   \hat f_{\rho\delta}(y) &= \frac{\pi^2}{m_l^3}\Big\{
      \hat y_\rho \hat y_\delta \; m_l|y| K_1(m_l|y|)
      + \delta_{\rho\delta} K_0(m_l|y|)
   \Big\},
   \\
   g_\rho(y) &= \frac{\pi^2}{m_l^2}\, y_\rho\, K_0(m_l|y|),
   \\
   q_\gamma(y) &= \frac{2\pi^2}{m_l^2}\,\hat y_\gamma\, K_1(m_l|y|),
   \\
   {f}_{\rho\delta\gamma}(y) &= \frac{\pi^2}{m_l^3}\Big\{
      \hat y_\gamma \hat y_\delta \hat y_\rho\, m_l|y|K_2(m_l|y|)
      + (
         \delta_{\rho\delta} \hat y_\gamma - \delta_{\gamma\rho} \hat y_\delta - 
         \delta_{\gamma\delta}\hat y_\rho
      ) \,K_1(m_l|y|)
   \Big\},
   \\
   p(|y|) &= \frac{2\pi^2}{m_l^2}\, K_0(m_l|y|).
\end{align}

\begin{figure}
   \includegraphics[width=0.49\textwidth,
      trim=0 0 0 30,clip]{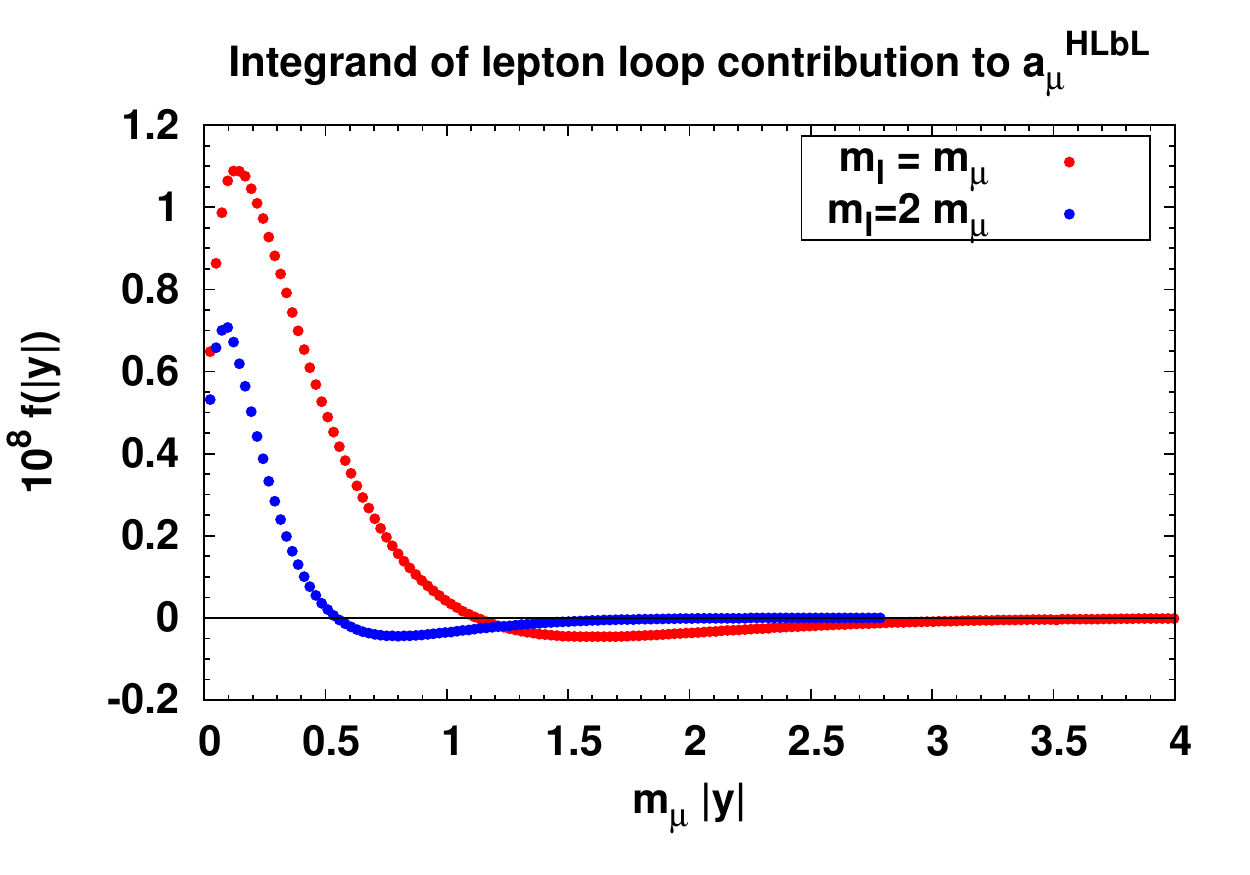}
   \includegraphics[width=0.49\textwidth,
      trim=0 0 0 30,clip]{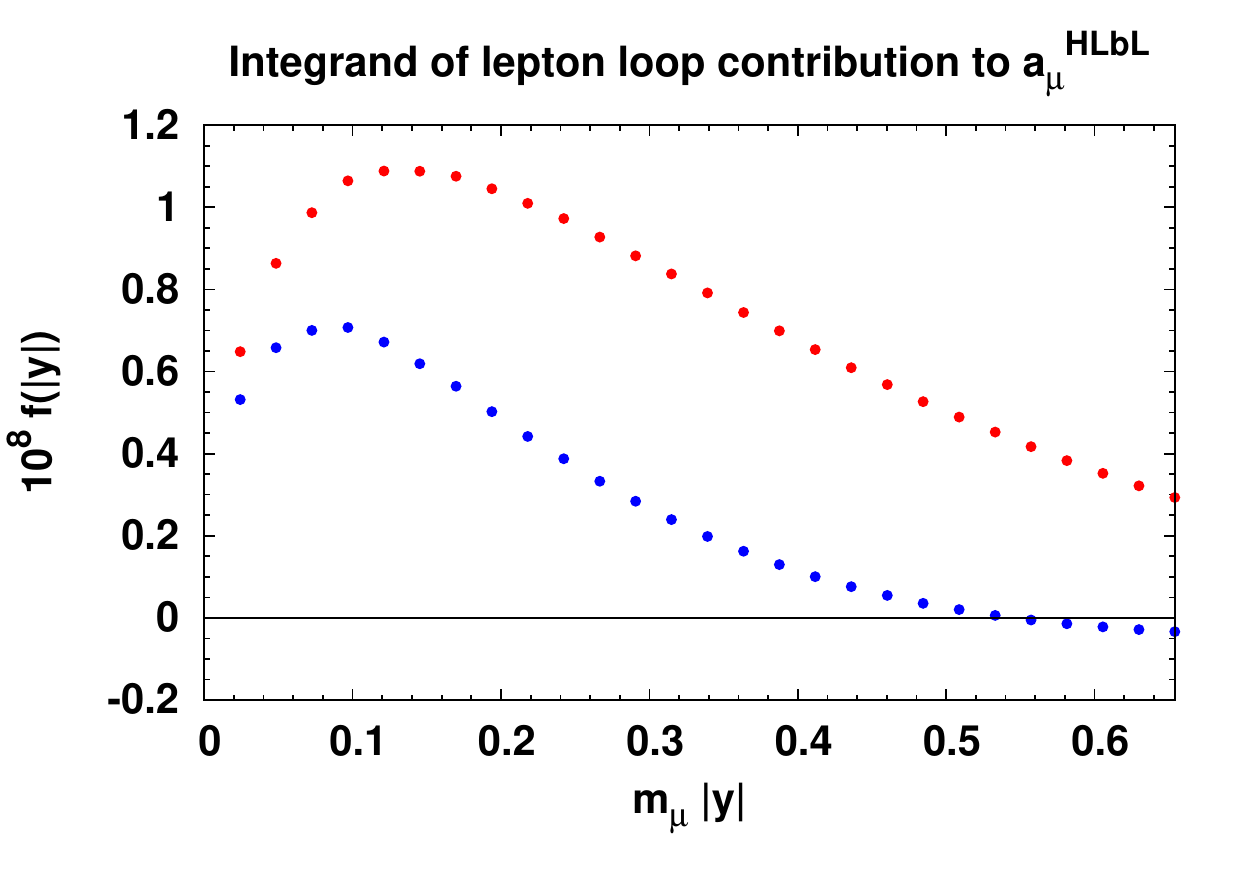}
   \caption{Integrand of the lepton loop contribution \(\amulbl\) in QED. The 
   full integration region is shown on the left, a detailed view of the small 
   $|y|$ region on the right.}
   \label{fig:leptonloop}
\end{figure}

The integrand \(f(\abs y)\) of the final $\abs y$ integration is shown in 
Fig.~\ref{fig:leptonloop}. The behaviour for small $\abs y$ is numerically 
compatible with $f(\abs{y})\propto m_\mu\abs y\log^2(m_\mu\abs y)$. This is 
quite steep and means that we probe the kernel precisely also for small 
distances. With this correlation function, the resulting value for $\amulbl$ for 
different loop masses can be reproduced at the percent level; see Table 
\ref{tab:leptonloop}.

\begin{table}
   \begin{center}
      \begin{tabular}{cSr@{}lSS}
         \toprule
           $m_l / m_\mu$
         & {$a_\mu^{\text{LbL}} \times 10^{11}$ (exact)}
         & \multicolumn{2}{c}{$a_\mu^{\text{LbL}} \times 10^{11}$}
         & {Precision}
         & {Deviation}
         \\ \midrule
         1   &  464.97 &  470.6&(2.3)(2.1) & 0.7\% & 1.2\% \\
         2   &  150.31 &  150.4&(0.7)(1.7) & 1.2\% & 0.06\%\\
         \bottomrule
      \end{tabular}
   \end{center}
   \caption{Results for the lepton-loop contribution in QED. For the exact 
     numbers cf.~\cite{Laporta:1992pa,Passera:PC}. The~first uncertainty stems 
     from the three-dimensional integration, the second from the extrapolation 
     to small $\abs y$.}
   \label{tab:leptonloop}
\end{table}

\section{Conclusions}
The covariant position-space method remains a promising approach to calculate the HLbL contribution to $(g-2)_\mu$. We did two 
tests of our QED kernel with the help of semi-analytic computations of the 
correlation function $i\hat\Pi$. The first test is the $\pi^0$-pole contribution 
in a vector-meson dominance model for the transition form factor, and the second 
is a lepton loop. We reproduce the known analytic result for the lepton loop at 
the percent level.
One important observation is that the $\pi^0$-pole contribution is very 
long-range, but we hope to be able to correct for the finite-size effects on 
this contribution, by computing the transition form 
factor~\cite{Gerardin:2016cqj} on the same ensemble and using 
Eqs.~(\ref{eq:PihatPi0Master},\,\ref{eq:PiK}).  We plan to make the QED kernel 
publicly available.

%\bibliography{references.bib}

\begin{thebibliography}{19}

\bibitem{Jegerlehner:2017lbd}
F.~Jegerlehner, \texttt{1705.00263}

\bibitem{hertzog:2015jru}
D.W. Hertzog, EPJ Web Conf. \textbf{118}, 01015 (2016), \texttt{1512.00928}

\bibitem{Blum:2014oka}
T.~Blum, S.~Chowdhury, M.~Hayakawa, T.~Izubuchi, Phys. Rev. Lett. \textbf{114},
  012001 (2015), \texttt{1407.2923}

\bibitem{Blum:2015gfa}
T.~Blum, N.~Christ, M.~Hayakawa, T.~Izubuchi, L.~Jin, C.~Lehner, Phys. Rev.
  \textbf{D93}, 014503 (2016), \texttt{1510.07100}

\bibitem{Blum:2016lnc}
T.~Blum, N.~Christ, M.~Hayakawa, T.~Izubuchi, L.~Jin, C.~Jung, C.~Lehner, Phys.
  Rev. Lett. \textbf{118}, 022005 (2017), \texttt{1610.04603}

\bibitem{Blum:2017cer}
T.~Blum, N.~Christ, M.~Hayakawa, T.~Izubuchi, L.~Jin, C.~Jung, C.~Lehner, Phys.
  Rev. \textbf{D96}, 034515 (2017), \texttt{1705.01067}

\bibitem{Asmussen:2015}
N.~Asmussen, J.~Green, V.~G\"ulpers, G.~von Hippel, H.B. Meyer, A.~Nyffeler,
  H.~Wittig, \emph{Hadronic light-by-light contribution to the muon anomalous
  magnetic moment on the lattice} (2015),
  \urlstyle{tt}\url{http://www.dpg-verhandlungen.de/year/2015/conference/heidelberg/part/sydm/session/3/contribution/6}

\bibitem{Green:2015mva}
J.~Green, N.~Asmussen, O.~Gryniuk, G.~von Hippel, H.B. Meyer, A.~Nyffeler,
  V.~Pascalutsa, PoS \textbf{LATTICE2015}, 109 (2016), \texttt{1510.08384}

\bibitem{Asmussen:2016lse}
N.~Asmussen, J.~Green, H.B. Meyer, A.~Nyffeler, PoS \textbf{LATTICE2016}, 164
  (2016), \texttt{1609.08454}

\bibitem{Colangelo:2014dfa}
G.~Colangelo, M.~Hoferichter, M.~Procura, P.~Stoffer, JHEP \textbf{09}, 091
  (2014), \texttt{1402.7081}

\bibitem{Colangelo:2014pva}
G.~Colangelo, M.~Hoferichter, B.~Kubis, M.~Procura, P.~Stoffer, Phys. Lett.
  \textbf{B738}, 6 (2014), \texttt{1408.2517}

\bibitem{Colangelo:2015ama}
G.~Colangelo, M.~Hoferichter, M.~Procura, P.~Stoffer, JHEP \textbf{09}, 074
  (2015), \texttt{1506.01386}

\bibitem{Colangelo:2017qdm}
G.~Colangelo, M.~Hoferichter, M.~Procura, P.~Stoffer, Phys. Rev. Lett.
  \textbf{118}, 232001 (2017), \texttt{1701.06554}

\bibitem{Colangelo:2017fiz}
G.~Colangelo, M.~Hoferichter, M.~Procura, P.~Stoffer, JHEP \textbf{04}, 161
  (2017), \texttt{1702.07347}

\bibitem{Pauk:2014rfa}
V.~Pauk, M.~Vanderhaeghen, Phys. Rev. \textbf{D90}, 113012 (2014),
  \texttt{1409.0819}

\bibitem{Gerardin:2016cqj}
A.~G\'erardin, H.B. Meyer, A.~Nyffeler, Phys. Rev. \textbf{D94}, 074507 (2016),
  \texttt{1607.08174}

\bibitem{Green:2015sra}
J.~Green, O.~Gryniuk, G.~von Hippel, H.B. Meyer, V.~Pascalutsa, Phys. Rev.
  Lett. \textbf{115}, 222003 (2015), \texttt{1507.01577}

\bibitem{Laporta:1992pa}
S.~Laporta, E.~Remiddi, Phys. Lett. \textbf{B301}, 440 (1993)

\bibitem{Passera:PC}
M.~Passera, private communication

\end{thebibliography}

\end{document}